\documentclass[12pt,amssymb]{article}
\usepackage{amssymb}
\usepackage{graphicx}

\def\gappeq{\mathrel{\rlap {\raise.5ex\hbox{$>$}}
{\lower.5ex\hbox{$\sim$}}}}

\def\lappeq{\mathrel{\rlap{\raise.5ex\hbox{$<$}}
{\lower.5ex\hbox{$\sim$}}}}

\def\Toprel#1\over#2{\mathrel{\mathop{#2}\limits^{#1}}}

\begin{document}
\pagestyle{empty}
\begin{flushright}
hep-ph/0302101\\
\end{flushright}
\vspace*{3mm}
\begin{center}
{\Large \bf  
The constraints on the non-singlet polarised parton densities 
from the infrared-renormalon model}\\
\vspace{0.1cm}
{\bf A.L. Kataev}\\
\vspace{0.1cm}
Institute for Nuclear Research of the Academy of Sciences of 
Russia,\\ 117312, Moscow, Russia\\
\end{center}
\begin{center}
{\bf ABSTRACT}\\
\end{center}

Using  the infrared-renormalon approach, 
we obtain  the constraints on the next-to-leading order non-singlet 
polarised parton densities. The advocated feature follows from the 
consideration  of the effect revealed in the process 
of the next-to-leading order fits to the  data for the assymetry 
of  polarised
lepton-nucleon scattering   
which result in the 
approximate nullification 
of the   $1/Q^2$-correction to $A_1^N(x,Q^2)$.

\noindent

The study of the QCD predictions for the photon-nucleon asymmetry 
$A_1^N=\frac{\sigma_{1/2}-\sigma_{3/2}}{\sigma_{1/2}+\sigma_{3/2}}$,
where subscripts denote the total angular momentum of the 
photon-nucleon pair along the incoming lepton's direction,
plays the essential role in the analysis of polarised deep inelastic scattering 
(DIS) (see e.g. Ref. \cite{Altarelli:1998gn}). It is related to the well-known 
structure function $g_1^N$ of polarised DIS by the following way 
\begin{equation}
\label{E1}
A_1^N(x,Q^2)=(1+\gamma^2)\frac{g_1^N(x,Q^2)}{F_1^N(x,Q^2)}
\end{equation}
where the kinematic factor $\gamma$ is defined as $\gamma=(4M_N^2x^2/Q^2)$ 
and $g_1^N(x,Q^2)$ is the structure function (SF) of 
polarised DIS, while $F_1^N(x,Q^2)$ SF 
enters into the cross-section of unpolarised charged lepton-hadron DIS 
(see e.g. Ref. \cite{Leader:2002ni}). Quite recently several procedures of the 
study of  the     
 $Q^2$-behaviour of $A_1^N$ were discussed in the literature (see Refs. 
\cite{Kotikov:1996vr}-\cite{Leader:2001kh}). 
Moreover, in Refs. \cite{Leader:1999qp,Leader:2001kh,Leader:2002ni} 
by fitting existing   CERN,  DESY and SLAC  data for polarised DIS 
in the kinematic region $0.005\leq x\leq 0.75$ and 
1 GeV$^2$$\leq Q^2\leq 58$ GeV$^2$ the  $1/Q^2$ dynamical power correction 
to $A_1^N$
was  extracted. 
In general it   gives additional contribution to the perturbation theory 
part of $(A_1^N)_{PT}$ and can  be parameterised as 
\begin{equation} 
\label{E2}
A_1^N(x,Q^2)=(A_1^N(x,Q^2))_{PT}+\frac{h^{A_1}(x)}{Q^2}~~~~.
\end{equation}
It is interesting, that in the process of the fits of 
Refs. \cite{Leader:1999qp,Leader:2001kh,Leader:2002ni} 
it was found that the $x$-shape of $h^{A_1}(x)$ is   consistent 
with zero (see e.g.  Fig.1 from Ref.\cite{Leader:2002ni}).

\begin{figure}
\includegraphics[width=8cm]{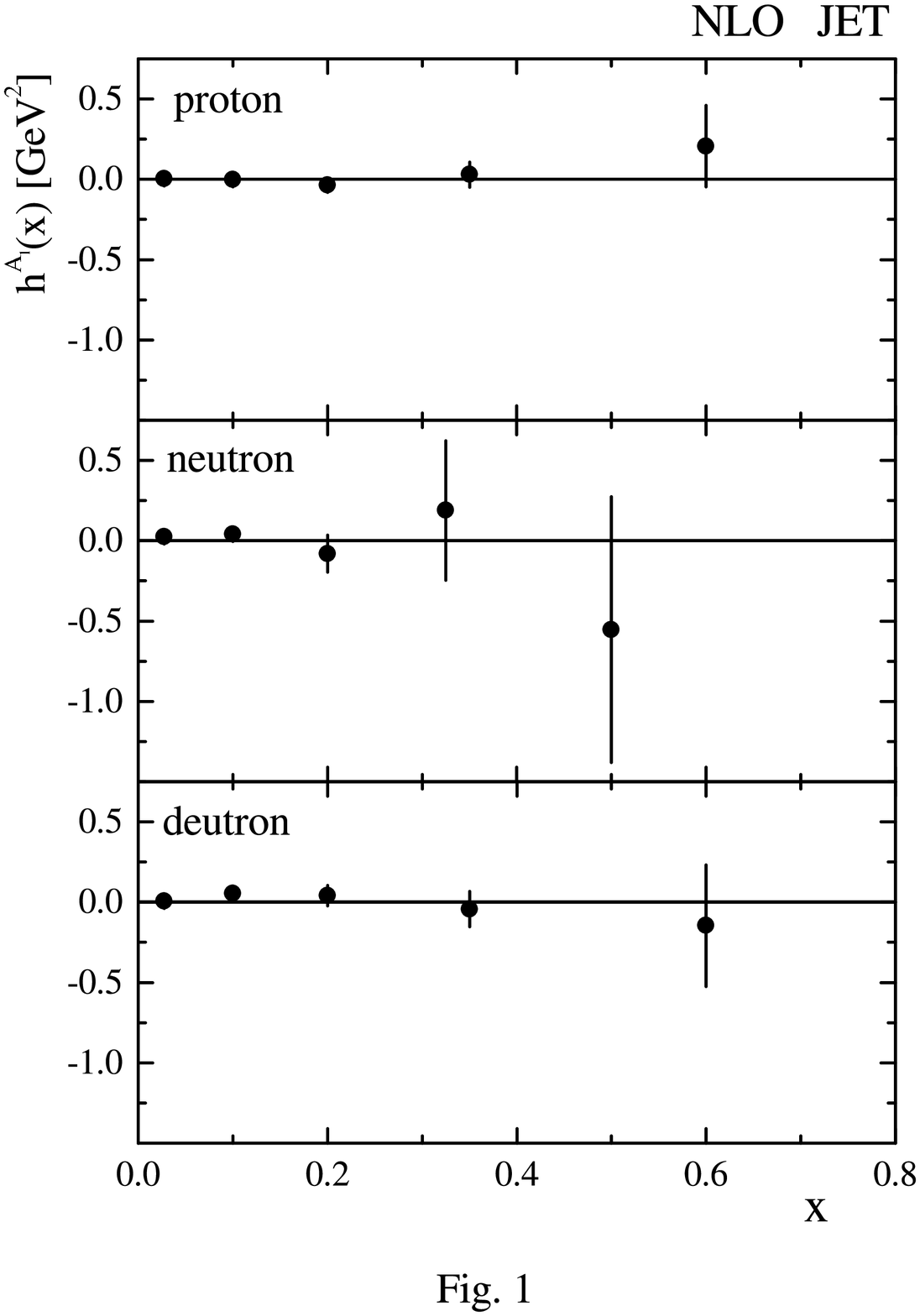}
\caption{The results of extraction of $h^{A_1}(x)$ from the next-to-leading 
order  fits of 
Ref. \cite{Leader:2002ni} in the JET scheme \cite{Carlitz:ab} .}   
\end{figure}

In this note we are describing the possible consequences 
of this effect in the 
non-singlet (NS) approximation, which is valid for the  $x$-cut $x\gtrsim 0.25$.
Our consideration will be based on  the infrared-renormalon (IRR) approach, 
developed  in QCD in Ref. \cite{Zakharov:1992bx} and reviewed in detail 
in Ref. \cite{Beneke:1998ui}.    
This 
approach was used in Ref. \cite{Dasgupta:1996hh} to 
study   the behaviour of the  $1/Q^2$ corrections 
to the NS contributions to  $F_2$ and $F_1$
SFs of unpolarised  DIS of charged leptons on nucleons 
and the pure NS  $xF_3$ SF of $\nu N$ DIS using 
$\overline{MS}$-scheme 
calculations\footnote{Note that we avoid considerations of the 
IRR renormalon free   expansions in QCD coupling constants with  
the ``freezing-type'' behaviour at small $Q^2$ (see 
Refs. \cite{Krasnikov:1996jq}).} .
It is interesting that  the predicted in Ref. \cite{Dasgupta:1996hh}
$x$-shape of the IRR induced power corrections to $xF_3$ was supported 
in Refs. \cite{Kataev:1997nc,Alekhin:1998df}
by the  leading order (LO) and next-to-leading order (NLO) 
fits to CCFR'97 data ( 
the detailed description and refinements of the fits of 
Ref. \cite{Kataev:1997nc} is given in 
Refs. \cite{Kataev:1999bp,Kataev:2001kk}). Therefore, it is 
worth to consider  the consequences  of calculations of the    
IRR 
contributions to the NS part of  $g_1^N$ SF  of polarised 
deep-inelastic scattering, which was also performed in 
Ref. \cite{Dasgupta:1996hh}.

Let us rewrite Eqs.(\ref{E1}),(\ref{E2}) in the following way
\begin{equation}
A_1^N=(1+\gamma^2)\frac{g_1^N(x,Q^2) (1+\frac{h^{g_1}(x)}{Q^2~g_1^N(x,Q^2)})}
{F_1^N(x,Q^2) (1+\frac{h^{F_1}(x)}{Q^2~F_1^N(x,Q^2)})}
\end{equation} 
where $h^{g_1}(x)/Q^2$ and $h^{F_1}(x)/Q^2$ are the model-independent 
parameterisations for the  twist-4 contributions to 
$g_1^N$ and $F_1^N$ SFs, which in general are non-zero. 
Using  the above mentioned effect of approximate 
nullification of the twist-4 correction to $A_1^N$ we get
\begin{equation}
\label{HT}
\frac{h^{g_1}(x)}{Q^2~g_1^N(x,Q^2)}\approx\frac{h^{F_1}(x)}{Q^2~F_1^N(x,Q^2)}~~~.
\end{equation} 
At the next step we will use  the existing inequality for $g_1^N(x,Q^2)$ 
SF, namely  
\begin{equation}
|g_1^N(x,Q^2)|\leq F_1^N(x,Q^2)~~.
\end{equation}
Combining it with  Eq. (\ref{HT}) we arrive to to the following bound 
\begin{equation}
\label{h}
|h^{g_1}(x)|\leq |h^{F_1}(x)| 
\end{equation}

It should be stressed that the calculations  of Ref. \cite{Dasgupta:1996hh}
predict that in the NS approximation (or in the valence-quarks approximation) 
the contributions 
of  the $1/Q^2$ corrections  to $F_1$ and $xF_3$
SFs are the same.
Indeed, the corresponding results of  Ref. \cite{Dasgupta:1996hh} can be 
re-written in the following way:
\begin{equation}
\label{F1}
h^{F_1}(x,\mu^2)=h^{F_3}(x,\mu^2)= A_2^{'}\int_x^1\frac{dz}{z}C_1(z)q^{NS}(x/z,\mu^2)
\end{equation} 
where 
\begin{equation}
\label{C1}
C_1(z)= -\frac{4}{(1-x)_{+}}+2(2+x+2x^2)-5\delta(1-x)-\delta^{'}(1-x)~~~,
\end{equation}
the $'+'$-prescription, for any test function, is defined as 
\begin{equation}
\int_0^1F(x)_{+}f(x)dx=\int_0^1F(x)[f(x)-f(1)]dx~~~,
\end{equation} 
and 
\begin{equation}
q^{NS}(x,\mu^2)=\sum_{i=1}^{n_f}\bigg(e_i^2-\frac{1}{n_f}\sum_{k=1}^{n_f}e_k^2
\bigg)
\bigg(q_i(x,\mu^2)+\overline{q}_i(x,\mu^2)\bigg)
\end{equation} 
are the  NS parton densities, 
$\mu^2$ is the normalisation point of order 1 GeV$^2$ and $A_2^{'}$ is  
is the IRR model parameter, to be extracted from the fits to the 
concrete data. Its value was extracted from the  low-energy 
$xF_3$ data, collected by the IHEP-JINR Neutrino detector at the 
IHEP 70 GeV proton synchrotron  \cite{Alekhin:2001zj}.
(see also Ref. \cite{Kataev:2001fv} for a review). The result of 
Ref. \cite{Alekhin:2001zj} $A_2^{'}$=$-0.10\pm 0.09$ (exp) GeV$^2$ is 
in agreement with 
the value extracted from the NLO analysis of the CCFR'97 $xF_3$ 
data \cite{Seligman:mc},
namely with  $A_2^{'}$=$-0.125\pm0.053$ (stat) GeV$^2$ \cite{Kataev:2001kk}. 
It should be noted, that the identity of Eq. (\ref{F1}) does not contradict 
point of view, expressed in Refs. \cite{Kotikov:1996vr,Kotikov:1998ew}, 
that to study the $Q^2$ behaviour of $A_1(Q^2)$ in the NS approximation 
it might be convenient to use the concrete $xF_3$ data  instead of theoretical expression for  $F_1^N$.

Consider  now the case of $g_1^N$ SF of polarised DIS. In general, the 
IRR contributions to $g_1^N$ were studied in Ref. \cite{Stein:1998wr}.
In the NS approximation the IRR contributions 
to $g_1^N$ were calculated in  Ref. \cite{Dasgupta:1996hh}, where the following 
result was obtained 
\begin{equation}
\label{g1}
h^{g_1}(x,\mu^2)= A_2\int_x^1\frac{dz}{z}C_1(z)\Delta^{NS}(x/z,\mu^2)~~~~.
\end{equation}
Here  $ \Delta^{NS}(x,\mu^2)$ are the NS polarised parton densities, namely 
\begin{equation}
\Delta^{NS}(x,\mu^2)=\sum_{i=1}^{n_f}\bigg(e_i^2-
\frac{1}{n_f}\sum_{k=1}^{n_f}e_k^2\bigg)
\bigg(\Delta q_i(x,\mu^2)+\Delta \overline{q}(x,\mu^2)\bigg)
\end{equation} 
and   the 
IRR model coefficient function $C_1(z)$ is the same,      
as in the case of the 
IRR model contributions to the  $1/Q^2$ corrections 
for $F_1$ and $xF_3$ SFs of unpolarised deep-inelastic scattering
(see Eq. (\ref{C1})). As to the IRR model parameter $A_2$, in general 
one should not expect that it has the same value as the parameter 
$A_2^{'}$ in Eq. (\ref{F1}). In principle, it should be extracted 
from the separate  fits to $g_1$ data in the NS approximation.
However, it is worth to stress, that in the NS approximation 
the IRR contributions to $g_1^N$ and $F_1^N$ are closely related
(the similar feature was revealed while 
comparing IRR model contributions to the Bjorken  sum rule for $g_1^N$
SF \cite{Broadhurst:1993ru} and still unmeasured  Bjorken sum rule for 
$F_1^{\nu N}$ SF \cite{Broadhurst:2002bi}).

Using now Eqs. (\ref{h}),(\ref{F1}) and Eq.(11), we get the following 
constraint 
\begin{equation}
|A_2 \Delta^{NS}(x,\mu^2)|\leq | A_2^{'} q^{NS}(x,\mu^2)|
\end{equation} 
which is valid both at the LO and NLO.
This constraint is the main result of our note.
The consequences for its $Q^2$-dependence can be further studied using 
the machinery of the DGLAP equations \

It is rather impressive that the NLO constraint of Eq. (13) is similar
to  the well-known LO bound of Ref. \cite{Altarelli:1998gn}, namely
\begin{equation}
|\Delta(x,Q^2)|\leq q(x,Q^2)~~~~~.
\end{equation}
Moreover, Eq. (13) can be also transformed to the LO relation between  the IRR model parameters
of Eq. (3) and Eq. (7), namely 
\begin{equation}
|A_2|\sim |A_2^{'}|~~.
\end{equation}
We hope that it will be possible to check  the 
relation of Eq. (15)  using 
the fits of the concrete  data for $g_1$ SF.

The final version of this work was presented at COMPASS meeting at Dubna 
(5 March 2003). I am grateful to M. G. Sapozhnikov for invitation.
It is the pleasure to thank  A. V. Sidorov and D.B. Stamenov for discussions
of the preliminary version of this note and to  G. Altarelli and B. Webber 
for the useful questions and comments.

This work is done within the program of the RFBR Grants N 02-01-00601 and 
N 03-02-17177.


\begin{thebibliography}{99}
\bibitem{Altarelli:1998gn}
G.~Altarelli, S.~Forte and G.~Ridolfi,
Nucl.\ Phys.\ B {\bf 534} (1998) 277
[hep-ph/9806345].
\bibitem{Leader:2002ni}
E.~Leader, A.~V.~Sidorov and D.~B.~Stamenov,
hep-ph/0212085, to be published in Phys.\ Rev.\ D.
\bibitem{Kotikov:1996vr}
A.~V.~Kotikov and D.~V.~Peshekhonov,
JETP Lett.\  {\bf 65} (1997) 7
[hep-ph/9612319].
\bibitem{Kotikov:1998ew}
A.~V.~Kotikov and D.~V.~Peshekhonov,
Eur.\ Phys.\ J.\ C {\bf 9} (1999) 55
[hep-ph/9810224].
\bibitem{Leader:1999qp}
E.~Leader, A.~V.~Sidorov and D.~B.~Stamenov,
in Proc. of   9th Lomonosov Conference on Elementary Particle Physics: 
Particle Physics on the Boundary of the Millenium, Moscow, Russia, 20-
26 Sep 1999, ed. A.I. Studenikin, World Scientific, 2001, p. 78.
\bibitem{Leader:2001kh}
E.~Leader, A.~V.~Sidorov and D.~B.~Stamenov,
Eur.\ Phys.\ J.\ C {\bf 23} (2002) 479
[hep-ph/0111267].
\bibitem{Carlitz:ab}
R.~D.~Carlitz, J.~C.~Collins and A.~H.~Mueller,
Phys.\ Lett.\ B {\bf 214} (1988) 229;
M.~Anselmino, A.~Efremov and E.~Leader,
Phys.\ Rept.\  {\bf 261} (1995) 1
[Erratum-ibid.\  {\bf 281} (1997) 399]
[hep-ph/9501369];\\
D.~Muller and O.~V.~Teryaev,
Phys.\ Rev.\ D {\bf 56} (1997) 2607
[hep-ph/9701413].
\bibitem{Zakharov:1992bx}
V.~I.~Zakharov,
Nucl.\ Phys.\ B {\bf 385} (1992) 452.
\bibitem{Beneke:1998ui}
M.~Beneke,
Phys.\ Rept.\  {\bf 317} (1999) 1
[hep-ph/9807443].
\bibitem{Dasgupta:1996hh}
M.~Dasgupta and B.~R.~Webber,
Phys.\ Lett.\ B {\bf 382} (1996) 273
[hep-ph/9604388].
\bibitem{Krasnikov:1996jq}
N.~V.~Krasnikov and A.~A.~Pivovarov,
Mod.\ Phys.\ Lett.\ A {\bf 11} (1996) 835
[hep-ph/9602272];\\
D.~V.~Shirkov and I.~L.~Solovtsov,
Phys.\ Rev.\ Lett.\  {\bf 79} (1997) 1209
[hep-ph/9704333];\\
Y.~A.~Simonov,
Phys.\ Atom.\ Nucl.\  {\bf 65} (2002) 135
[Yad.\ Fiz.\  {\bf 65} (2002) 140]
[hep-ph/0109081].
\bibitem{Kataev:1997nc}
A.~L.~Kataev, A.~V.~Kotikov, G.~Parente and A.~V.~Sidorov,
Phys.\ Lett.\ B {\bf 417} (1998) 374
[hep-ph/9706534].
\bibitem{Alekhin:1998df}
S.~I.~Alekhin and A.~L.~Kataev,
Phys.\ Lett.\ B {\bf 452} (1999) 402
[hep-ph/9812348].
\bibitem{Kataev:1999bp}
A.~L.~Kataev, G.~Parente and A.~V.~Sidorov,
Nucl.\ Phys.\ B {\bf 573} (2000) 405
[hep-ph/9905310].
\bibitem{Kataev:2001kk}
A.~L.~Kataev, G.~Parente and A.~V.~Sidorov,
Fiz.\ Elem.\ Chast.\ Atom.\ Yadra {\bf 34} (2003) 43
[Phys.\ Part.\ Nucl.\  {\bf 34} (2003) 20]
[hep-ph/0106221].
\bibitem{Alekhin:2001zj}
S.~I.~Alekhin {\it et al.}, IHEP-JINR Neutrino Detector collaboration,
Phys.\ Lett.\ B {\bf 512} (2001) 25
[hep-ex/0104013].
\bibitem{Kataev:2001fv}
A.~L.~Kataev, hep-ph/0107247, in Proc. of 15th Les Recontre de 
Physique de la Vallee d'Aosta :
Results and Perspectives in Particle Physics, 2001,ed. M. Greco, Frascati 
Physics Series, v XXII, p.205.
\bibitem{Seligman:mc}
W.~G.~Seligman {\it et al.}, CCFR collaboration,
Phys.\ Rev.\ Lett.\  {\bf 79} (1997) 1213.
\bibitem{Stein:1998wr}
E.~Stein, M.~Maul, L.~Mankiewicz and A.~Schafer,
Nucl.\ Phys.\ B {\bf 536} (1998) 318
[hep-ph/9803342].
\bibitem{Broadhurst:1993ru}
D.~J.~Broadhurst and A.~L.~Kataev,
Phys.\ Lett.\ B {\bf 315} (1993) 179
[hep-ph/9308274].
\bibitem{Broadhurst:2002bi}
D.~J.~Broadhurst and A.~L.~Kataev,
Phys.\ Lett.\ B {\bf 544} (2002) 154
[hep-ph/0207261].
\bibitem{Gribov:ri}
V.~N.~Gribov and L.~N.~Lipatov,
Yad.\ Fiz.\  {\bf 15} (1972) 781
[Sov.\ J.\ Nucl.\ Phys.\  {\bf 15} (1972) 438];\\
G.~Altarelli and G.~Parisi,
Nucl.\ Phys.\ B {\bf 126} (1977) 298;\\
Y.~L.~Dokshitzer,
Sov.\ Phys.\ JETP {\bf 46} (1977) 641
[Zh.\ Eksp.\ Teor.\ Fiz.\  {\bf 73} (1977) 1216].
\end{thebibliography}
\end{document}